\newcommand{\ls}[1]     
{\dimen0=\fontdimen6\the=#1\dimen0
 \advance\lineskip.5\fontdimen5\the\lineskip-\dimen0
 \lineskiplimit=.9\lineskip
 \baselineskip=\lineskip     \advance\baselineskip\dimen0
 \normallineskip\lineskip
 \normallineskiplimit\lineskiplimit
 \normalbaselineskip\baselineskip
 \ignorespaces
}

\documentclass[a4paper]{article}
\usepackage{INTERSPEECH2020}

\usepackage[normalem]{ulem}
\usepackage{amssymb}
\usepackage{cite}
\usepackage{xcolor}
\usepackage{hyperref}
\usepackage{graphics}
\usepackage{epsfig}
\usepackage{graphicx}
\usepackage{subfigure}
\usepackage{textcomp}
\usepackage{multirow}
\usepackage{multicol}

\title{SERIL: Noise Adaptive Speech Enhancement using Regularization-based Incremental Learning}

\name{Chi-Chang Lee$^{1,2}$, Yu-Chen Lin$^{1,2}$, Hsuan-Tien Lin$^{1}$, Hsin-Min Wang$^{3}$, Yu Tsao$^{2}$}

\address{
\small{
{$^{1}$Department of Computer Science and Information Engineering, National Taiwan University, Taiwan\\
$^{2}$Research Center for Information Technology Innovation, Academia Sinica, Taiwan\\
$^{3}$Institute of Information Science, Academia Sinica, Taiwan}
}
}

\email{
r08922a27@csie.ntu.edu.tw,
f04922077@csie.ntu.edu.tw,
htlin@csie.ntu.edu.tw,
whm@iis.sinica.edu.tw,
yu.tsao@citi.sinica.edu.tw
}

\begin{document}
\maketitle
\begin{abstract}
Numerous noise adaptation techniques have been proposed to fine-tune deep-learning models in speech enhancement (SE) for mismatched noise environments.
Nevertheless, adaptation to a new environment may lead to catastrophic forgetting of the previously learned environments.
The catastrophic forgetting issue degrades the performance of SE in real-world embedded devices, which often revisit previous noise environments.
The nature of embedded devices does not allow solving the issue with additional storage of all pre-trained models or earlier training data.
In this paper, we propose a regularization-based incremental learning SE (SERIL) strategy, complementing existing noise adaptation strategies without using additional storage.
With a regularization constraint, the parameters are updated to the new noise environment while retaining the knowledge of the previous noise environments.
The experimental results show that, when faced with a new noise domain, the SERIL model outperforms the unadapted SE model.
Meanwhile, compared with the current adaptive technique based on fine-tuning, the SERIL model can reduce the forgetting of previous noise environments by 52\%.
The results verify that the SERIL model can effectively adjust itself to new noise environments while overcoming the catastrophic forgetting issue.
The results make SERIL a favorable choice for real-world SE applications, where the noise environment changes frequently.

\end{abstract}


\ls{0.9}

\section{Introduction}
\label{sec:intro}

The objective of speech enhancement (SE) is to transform low-quality speech signals into enhanced-quality speech signals \cite{SE}.
In many speech-related applications such as automatic speech recognition (ASR) \cite{ASR1} and speech emotion recognition \cite{emotion2}, SE is used as a preprocessor to remove noise components from speech signals.
In many portable or assistive-hearing devices, such as mobile phones \cite{tan2019real}, hearing aids \cite{aids}, and cochlear implants \cite{implants}, SE is crucial for increasing speech intelligibility and quality in noise environments.

In the past few years, deep learning (DL)-based models have been widely used for SE \cite{SE1, SE2, SE3, SE4, SE5, SE6, DAELu, CHLee, IA-NET}.
Various deep neural networks such as convolutional neural networks (CNNs), recurrent neural networks (RNNs), and long short-term memory (LSTM) have been used as fundamental models in SE systems.
In these systems, some metrics are defined to measure the distance between the enhanced output and the clean reference, and the DL models are trained to minimize the distance.
The \emph{L1} and \emph{L2} (mean-square-error) losses are commonly used because of their ease of computation and differentiability.
However, these two losses may not be optimal for specific tasks, and thus other metrics have been used as the loss to train the DL models \cite{metricgan, wang2018maximum}.

In addition to model types and loss functions, another important consideration for the success of an SE system is its ability to adapt to new environments, particularly when deployed in embedded devices.
In real-world situations, the noise in the testing environment is unseen in the training set; moreover, the noise types often vary over time.
The mismatch between training and testing environments can significantly degrade the performance of SE.
Therefore, identifying an approach that can efficiently and effectively adapt DL models to new testing conditions and improve the performance of SE is necessary.
Thus far, several domain adaptation approaches \cite{adaptation1, adaptation2, fine-tune1, fine-tune2} have been proposed to address the training-testing acoustic mismatch issue, which is also known as the \emph{domain shift} problem.
Although noise-adapted models can provide improved SE results for these conventional approaches, they often suffer from a \emph{catastrophic forgetting} effect\cite{forget1, forget2}.
In other words, when DL models adapt to a new noise environment, they usually perform poorly when dealing with previously adapted noise environments.


In this paper, we propose a regularization-based incremental learning strategy for adapting DL-based SE models to new environments (speakers and noise types) while handling the catastrophic forgetting issue. The proposed method is termed SERIL.
SERIL exploits the advantages of two well-known incremental learning algorithms: (1) whole past optimization path information\cite{SI} and (2) curvature-based strategy\cite{EWC}.
We evaluated SERIL using two datasets: the Voice Corpus Bank corpus (VCB) \cite{VCB} and the TIMIT corpus \cite{TIMIT}, which were used to form the training and testing sets, respectively.
The overall SERIL included two phases: offline and online.
In the offline phase, we first trained the DL model on the utterances from the VCB corpus with 13 different types of noise.
In the online phase, SERIL first adapted the pre-trained model based on a small amount of adaptation data; then, the adapted model was used for SE.
A direct fine-tuning model adaptation approach was implemented for comparison.
Experimental results show that SERIL and the direct fine-tuning approach both effectively adapt the SE model to new environments and improve SE performance, compared with the pre-trained DL model without adaptation.
Moreover, compared to the direct fine-tuning approach, SERIL maintained high SE performance against all previously learned types of noise, thus effectively addressing the catastrophic forgetting problem.

The remainder of this paper is organized as follows.
Section \ref{sec:motivation} presents some related work and explains the motivation of using incremental learning strategy to help noise adaptation issue on speech enhancement.
In Section \ref{sec:method}, we detail the philosophies of the proposed SERIL system.
The experimental setup and results are then reported in Section \ref{sec:evaluation}.
Finally, Section \ref{sec:conclusion} presents some concluding remarks.

\section{Related Work and Motivation}
\label{sec:motivation}

An intuitive SE method to overcome the mismatch problem is to collect as many types of noise as possible to increase the generalization ability\cite{CHLee}.
However, it is impractical to cover the infinite types of noise that may be encountered in real situations.
Several researches \cite{fine-tune1, fine-tune2} have been proposed to directly fine-tune a pre-trained model to improve the performance in a target domain.
When entering a new circumstance, these algorithms only focus on the current noise domain, and ignore the memory of the previously learned noise types.
In many applications, such as edge-devices, the type of noise changes frequently, and it is common to re-encounter learned types of noise.
However, the adapted SE model cannot perform well in the previously learned noise types. This effect is called catastrophic forgetting \cite{forget1, forget2}.
Although the SE model can be fine-tuned every time the environment is changed, the repeated model adaptation process will result in high computation and time costs.

\vspace{-0.5em}
\begin{figure}[h]
\centering
\includegraphics[width=0.75\columnwidth]{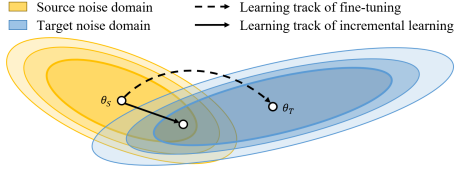}
\vspace{-0.5em}
\caption{Relationship between fine-tuning and incremental learning from source noise domain to unseen target domain.}
\vspace{-0.5em}
\label{fig:2}
\end{figure}

The above limitations of adaptive methods based on direct fine-tuning motivated us to apply the incremental learning algorithm to SE.  
Incremental learning is also known as \emph{continuous learning} or \emph{life-long learning}.
Figure \ref{fig:2} illustrates the relationship between direct fine-tuning and incremental learning.
Training trajectories are illustrated in a schematic parameter space, with parameter regions leading to good performance on the source (yellow region) and target (blue region), denoted as tasks $S$ and $T$, respectively.
After learning in task $S$, the parameters are located in $\theta_{S}$.
As shown by the dashed arrow in Figure~\ref{fig:2}, when the SE model is adapted by taking gradient steps to minimize the loss based on task $T$ alone, the resulting $\theta_{T}$ is beyond the good performance area of Task $S$, i.e., what is already learned in Task $S$ is forgotten.
In contrast, in incremental learning, the SE model weights are updated to the target domain while retaining the knowledge learned from the source domain.
This is often realized by finding the overlapping region of the source and target domains. The learning trajectory of incremental learning shown by the solid arrow in Figure \ref{fig:2} illustrates this concept.
In this way, incremental learning can help the resulting model provide good SE results in the target domain while maintaining satisfactory performance in the source domain.


\section{The SERIL System}
\label{sec:method}



\subsection{Architecture and loss function of the SERIL system}
\label{sec:overall}

\begin{figure}[h]
\centering
\includegraphics[width=0.95\columnwidth]{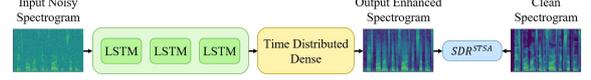}
\vspace{-0.5em}
\caption{Architecture of the SERIL system using the short-time spectral amplitude SDR ($SDR^{STSA}$) as the loss function.}
\vspace{-1.5em}
\label{fig:framework}
\end{figure}

The architecture of the SERIL system is depicted in Figure~\ref{fig:framework}.
The system performs SE in the spectral domain.
Speech waveforms are first converted into time-frequency features using a 512-point short-time Fourier transform (STFT) with a hamming window size of 32 ms and a hop size of 16 ms.
Each feature vector consists of 257 elements.
The enhanced spectral features are then converted into the waveform domain by inverse STFT with an overlap-add method.
In the SERIL system, the first 3 layers are LSTM layers (one-directional LSTM was used for achieving real-time inference). The hidden dimension of each LSTM is 257. A fully connected layer is concatenated to the output of the last LSTM layer for scaling. 


As mentioned earlier, the L1 and L2 norms are commonly used as the loss function to train DL-based SE models.
In this study, we derived another loss function based on the short-time spectral amplitude SDR  (\(SDR^{STSA}\)), which was shown to provide better results than L1 and L2 norms in our preliminary experiments.
In a previous study, Kolbæk et al. \cite{TSDR} reported that using the time-domain SDR \cite{SDR1, SDR2} as the loss can help the SE models to achieve improved performance.
Because the input and output of SERIL are both spectral features, we need to modify the original SDR loss to use it in the spectral domain.
We note that SDR can be regarded as the energy ratio of enhanced speech projected on the clean speech space over enhanced speech projected on the orthogonal space of clean speech.
By Parseval's theorem \cite{Parseval} and the linear property of Fourier transform, the energy ratio in the time domain is equivalent to that in the time-frequency domain.
Therefore, we define the (\(SDR^{STSA}\)) as follows:
\begin{equation}
{
    SDR^{STSA}(\hat X, X) = 10log_{10}\frac{\|\alpha X\|^{2}}{\|\alpha X - \hat X\|^{2}}.
}
\end{equation}
Given the noisy spectral features, $Y$, the SE model aims to generate enhanced spectral features, $\hat X$. $\alpha$ is computed by \((X\cdot\hat X)/\|X\|^{2}\), where $X$ is the target clean spectral features.
In addition, $f_{\theta}(.)$ is equal to $\hat X$; thus, we denote our loss function $-SDR^{STSA}(f_{\theta}(Y), X)$ as $l_{\theta}(Y)$.



\subsection{Curvature-based regularization strategy}

Considering the losses in the previous and new acoustic environments, \(L_{old}\) and \(L_{new}\), respectively, the total loss can be formulated as:
\begin{equation}
\label{eq:org}
{
    L(\theta)=L_{new}(\theta) + L_{old}(\theta).
}
\end{equation}
Because the training data of the previous environment is usually not accessible online, we cannot calculate $L_{old}(\theta)$.
Instead, we can assume that the loss of the previous environment can be revealed from the learned SE model, $\theta$.
By approximating \(L_{old}\) using the second-order Taylor expansion at \(\theta = \theta^{*}\), we have
\begin{equation}
{
    \label{eq:taylor1}
    L_{old}(\theta) \approx 
     L_{old}(\theta^{*}) + 
     {\delta \theta}^{T} 
    \nabla_{\theta} L_{old}(\theta^{*}) + \frac{1}{2}{\delta \theta}^{T} H(\theta^{*}) {\delta \theta}, 
}
\end{equation}
where \(\delta \theta\) is \(\theta - \theta^{*}\); \(H(\theta^{*})\) is the Hessian matrix of \(L_{old}\) at \(\theta = \theta^{*}\); and \(L_{old}(\theta^{*})\) is a constant.
Because the elements in \(\nabla_{\theta} L_{old}(\theta^{*})\) are generally small enough to be ignored, we can obtain the approximate form as \(L_{old}(\theta) \approx \frac{1}{2}{\delta \theta}^{T} H(\theta^{*}) {\delta \theta}\). 
Similar to the elastic weight consolidation \cite{EWC, Online-EWC}, we ignore the cross terms in $H(\theta^{*})$ to improve computational efficiency.
The approximate form becomes
\begin{equation}
{
    \label{eq:hessian}
    H(\theta^{*}) \approx diag( \mathbb{E}_{Y \sim D_{old}}[(\nabla_{\theta}l_{\theta}(Y))(\nabla_{\theta}l_{\theta}(Y))^{T}])|_{\theta = \theta^{*}},
}
\end{equation}
where \(Y\) is the speech sample from the previous environment \(D_{old}\).
Finally, substituting (\ref{eq:taylor1}) and (\ref{eq:hessian}) into (\ref{eq:org}), we have
\begin{equation}
{
    L(\theta) \approx L_{new}(\theta)+\lambda\sum_{i}F_{\theta_{i}}(\theta_{i}-\theta_{i}^{*})^2,
}
\end{equation}
where $\lambda$ is a hyperparameter; \(i\) is the index of the parameters in the model; $\theta_{i}$ and $\theta^{*}_{i}$ are the \(i\)-th parameters in the current and previous environments, respectively; and \(F_{\theta_{i}}\) is the diagonal element of \(H(\theta^{*})\).
The intuitive interpretation of \(F_{\theta_{i}}\) is the local curvature, which indicates the sensitivity that affects the performance of the previous acoustic environment.

Kolouri et al. \cite{SCP} provided a different explanation for the geometric view of the regularization term, which can be applied to our scenario.
As \(\theta \to \theta^{*}\), \(\frac{1}{2}\|\theta - \theta^{*}\|^{2}_{F_{\theta_{i}}}\) can be interpreted as the expectation of the squared difference of the loss values of the training samples of the previous environment, i.e., \(\mathbb{E}_{Y \sim D_{old}}[\frac{1}{2}(l_{\theta}(Y) - l_{\theta^{*}}(Y))^{2}]\).
Similar to (\ref{eq:taylor1}), the distance can be approximated by \( \sum_{i}F_{\theta_{i}}(\theta_{i}-\theta_{i}^{*})^2\), which is also derived by the second-order Taylor expansion of \(\mathbb{E}_{Y \sim D_{old}}[\frac{1}{2}(l_{\theta}(Y) - l_{\theta^{*}}(Y))^{2}]\) at \(\theta = \theta^{*}\).
Referring to \cite{Online-EWC,RWalk,SCP}, we apply the interpolation approach to the case of multiple tasks. Given \(\tilde F_{\theta}^{t-1}\) derived by all previous tasks, \(\tilde F_{\theta}^{t}\) is updated as
\begin{equation}
{
    \tilde F_{\theta}^{t} = \alpha F_{\theta}^{t} + (1 - \alpha ) \tilde F_{\theta}^{t-1},
}
\end{equation}
where \(t\) is the index of the task; $\alpha$ is a hyperparameter in [0,1]; \(F_{\theta}^{t}\) denotes \(F_{\theta}\) derived from the ($t-1$)-th task; and \(\tilde F_{\theta}^{t}\) is the interpolation result of \(\tilde F_{\theta}^{t-1}\) and \(F_{\theta}^{t}\), corresponding to the information of past accumulations and curvatures.

\subsection{Path optimization augmenting approach}

\begin{figure}[h]
    \centering
    \includegraphics[width=0.85\columnwidth]{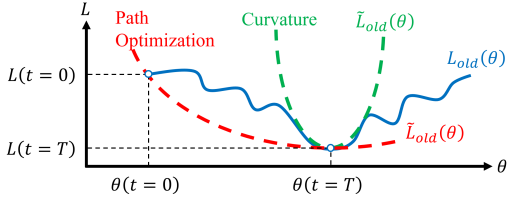}
    \caption{Relationship between the real loss (blue), curvature-based approximate loss (green), and path optimization-based approximate loss (red) while adapting the SE model. $t=0$ and $t=T$ are the start and end times, respectively.}
    \label{fig:curve}
\end{figure}

Although \(F_{\theta}\) is equipped with rationality to avoid catastrophic forgetting, the commonly used curvature-based methods \cite{EWC,Online-EWC} of deriving \(F_{\theta}\) rely on point estimation, which only capture local curvature information around \(\theta^{*}\).
In contrast, the path optimization-based method \cite{SI} considers the information over the optimization path on the loss surface.
In particular, the importance score is determined by accumulating over the entire training trajectory, as illustrated in Figure~\ref{fig:curve}.

By using the first-order Taylor approximation and setting \(t_{s}\) and \(t_{e}\) as the start and end steps of the $t$-th task, the change in loss \(L\) over the time from \(t_{s}\) to \(t_{e}\) can be written as
\begin{equation}{
    \begin{aligned}
        L(\theta(t_{e})) - L(\theta(t_{s})) & \approx \int_{t_{s}}^{t_{e}} (\nabla_{\theta}L(\theta(t))\cdot\frac{\theta(t)}{dt})dt \\
        & = \sum_{i}(\int_{t_{s}}^{t_{e}}    \frac{\partial L} {\partial\theta_{i}}\frac{d\theta_{i}}{dt}dt), 
    \end{aligned}}
\end{equation}
where \(i\) is the index of the SE model parameter.
To simplify the description, we denote \((\int_{t_{s}}^{t_{e}} \frac{\partial L} {\partial\theta_{i}}\frac{d\theta_{i}}{dt}dt)\) as \(-\Delta L_{i}^{t}\).
Therefore, the change in the total loss can be represented as the summation of the individual loss \(\Delta L_{i}^{t}\) associated with each parameter.
We put a minus sign on the left side of \(\Delta L_{i}^{t}\) to make the sign consistent with the regularization term.
Practically, we replace \(\int_{t_{s}}^{t_{e}} \frac{\partial L}{\partial\theta_{i}}\frac{d\theta_{i}}{dt}dt\) with \(\sum_{\tau = t_{s}}^{t_{e} - 1} \frac{\partial L} {\partial\theta_{i}}(\theta_{i}(\tau+1) - \theta_{i}(\tau))\), where \(\tau\) is the index of iteration.
From \cite{SI}, the definition of importance scores as we begin to train the \(t\)-th task can be defined as
\begin{equation}
{
    S_{\theta_{i}}^{t} = \sum_{t'<t} \frac{\Delta L_{i}^{t'}}{(\Delta \theta_{i}^{t'})^{2} + \epsilon},
}
\end{equation}
where $t'$ is the index of the task before the \(t\)-th task; $\theta_{i}^{t'}$ is the $i$-th parameter of the SE model derived from training the $t'$-th task; $\Delta \theta_{i}^{t'}$ is $\theta_{i}^{t'} - \theta_{i}^{t'-1}$; and $\epsilon$ is a hyperparameter with a positive value.

Similar to \cite{RWalk}, we combined the advantages of curvature-based\cite{EWC,Online-EWC} and path optimization-based\cite{SI} approaches.
The importance of parameter \(\theta_{i}\) when training the \(t\)-th task can be written as \(((1 - \beta)\tilde F_{\theta_{i}}^{t}+\beta S_{\theta_{i}}^{t})\).
Therefore, the training loss is defined as:
\begin{equation}
{
    \tilde L^{t}(\theta) = L^{t}(\theta) + \lambda \sum_{i} ((1 - \beta)\tilde F_{\theta_{i}}^{t}+\beta S_{\theta_{i}}^{t})(\theta_{i} - \theta_{i}^{t-1})^{2},
}
\end{equation}
where \(t\) is the index of the task (if $t$ is zero, \(\tilde L^{t}(\theta)\) is equivalent to \(L^{t}(\theta)\));
\(\theta_{i}^{t-1}\) is the \(i\)-th parameter after training the ($t-1$)-th task; and $\beta$ is a scalar with the value in [0,1], which determines the weight of the two strategies.

\begin{figure*}[ht]
    \centering
    
    \subfigure[$E_{0}$: original]{
        \includegraphics[width=0.17\linewidth]{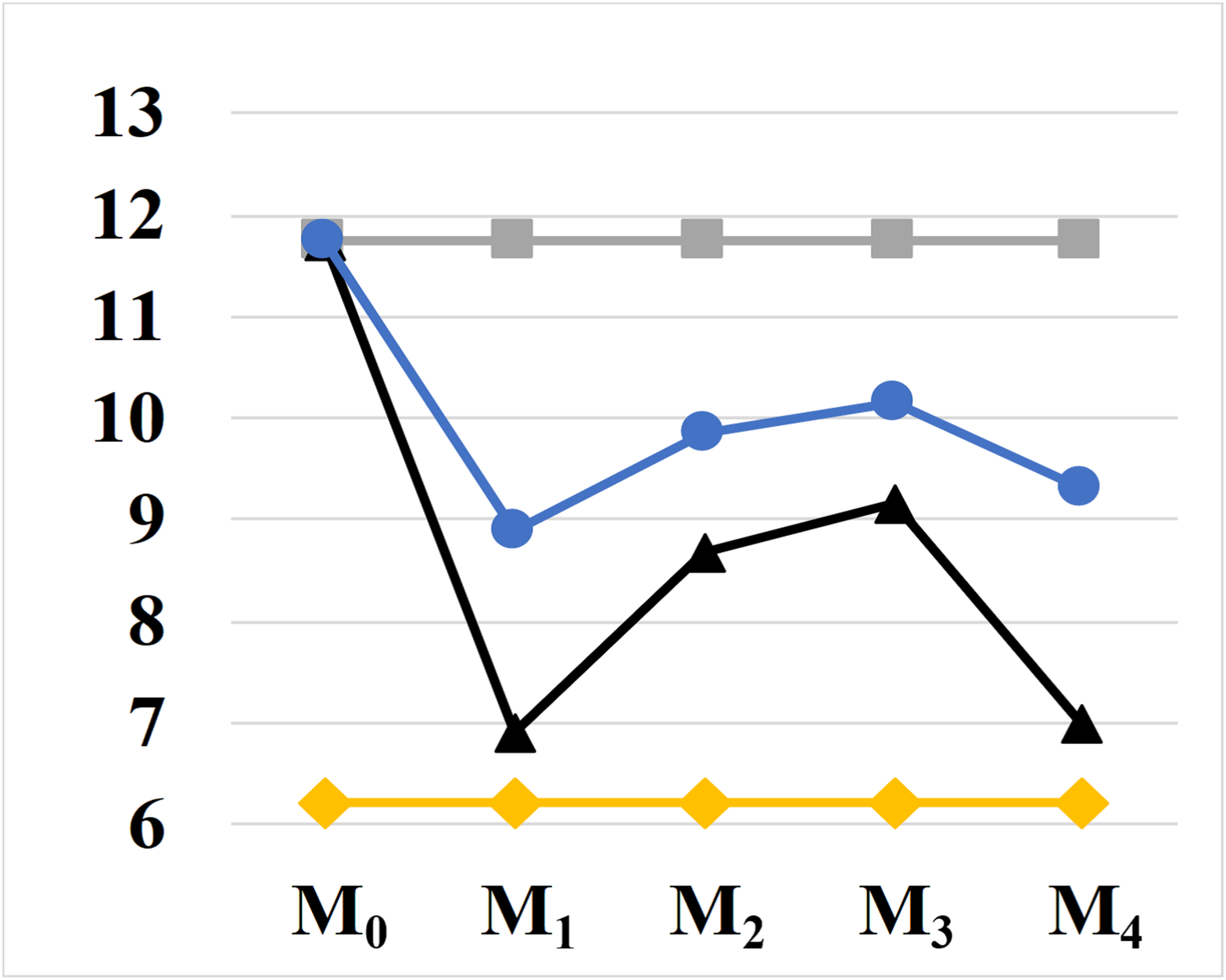}
    }
    \subfigure[$E_{1}$: cough]{
        \includegraphics[width=0.17\linewidth]{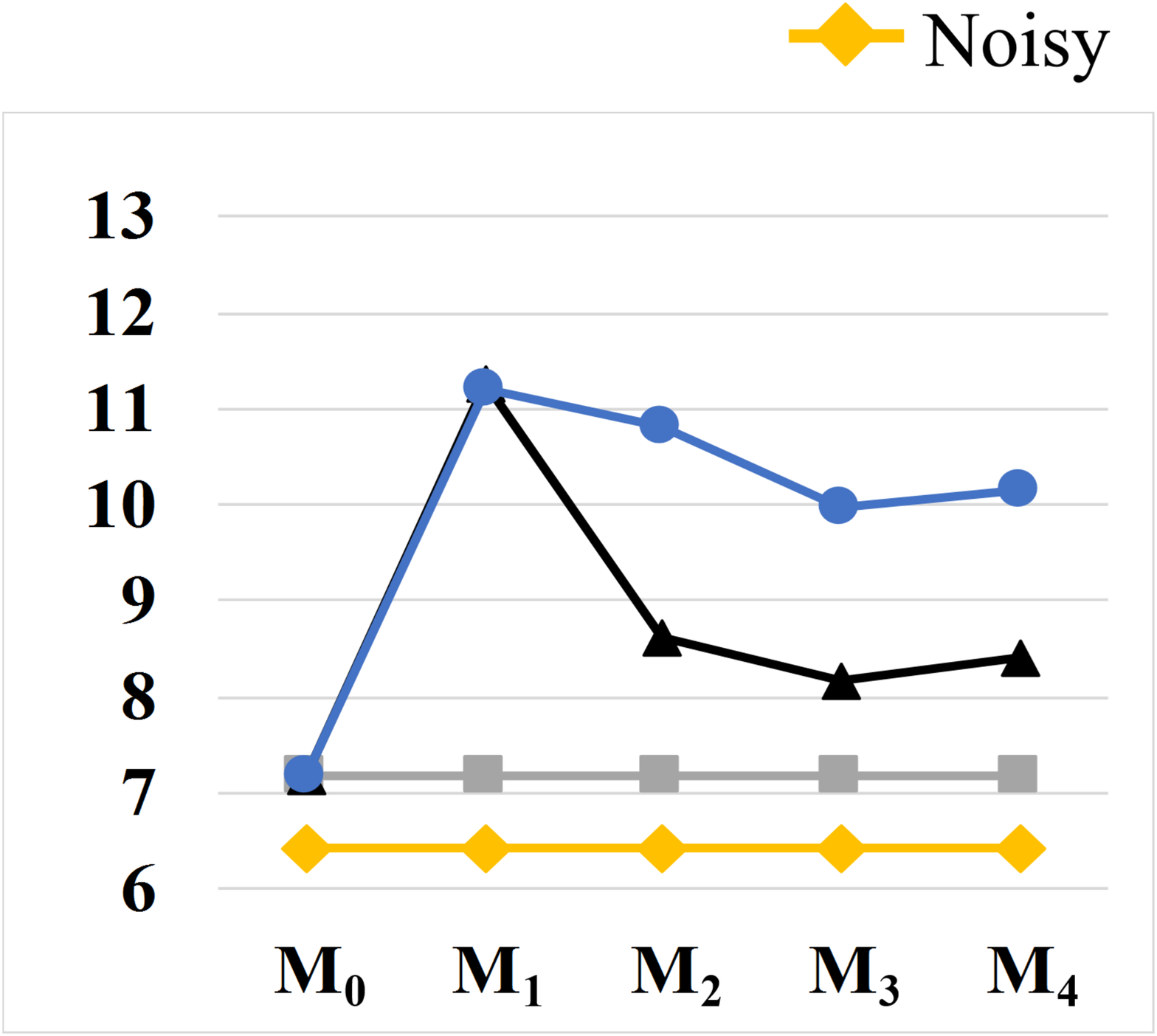}
    }
    \subfigure[$E_{2}$: door moving]{
        \includegraphics[width=0.17\linewidth]{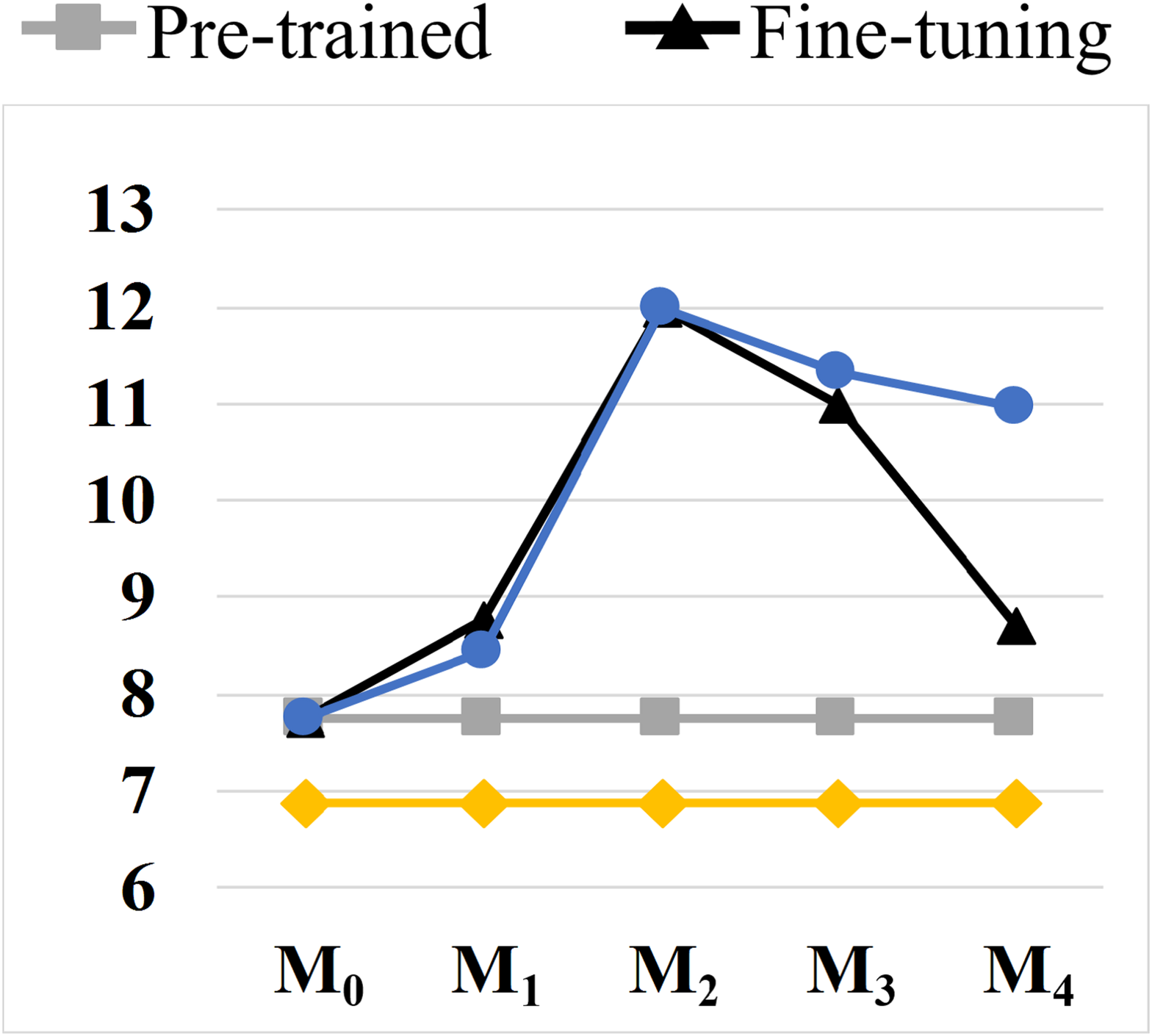}
    }
    \subfigure[$E_{3}$: footsteps]{
        \includegraphics[width=0.17\linewidth]{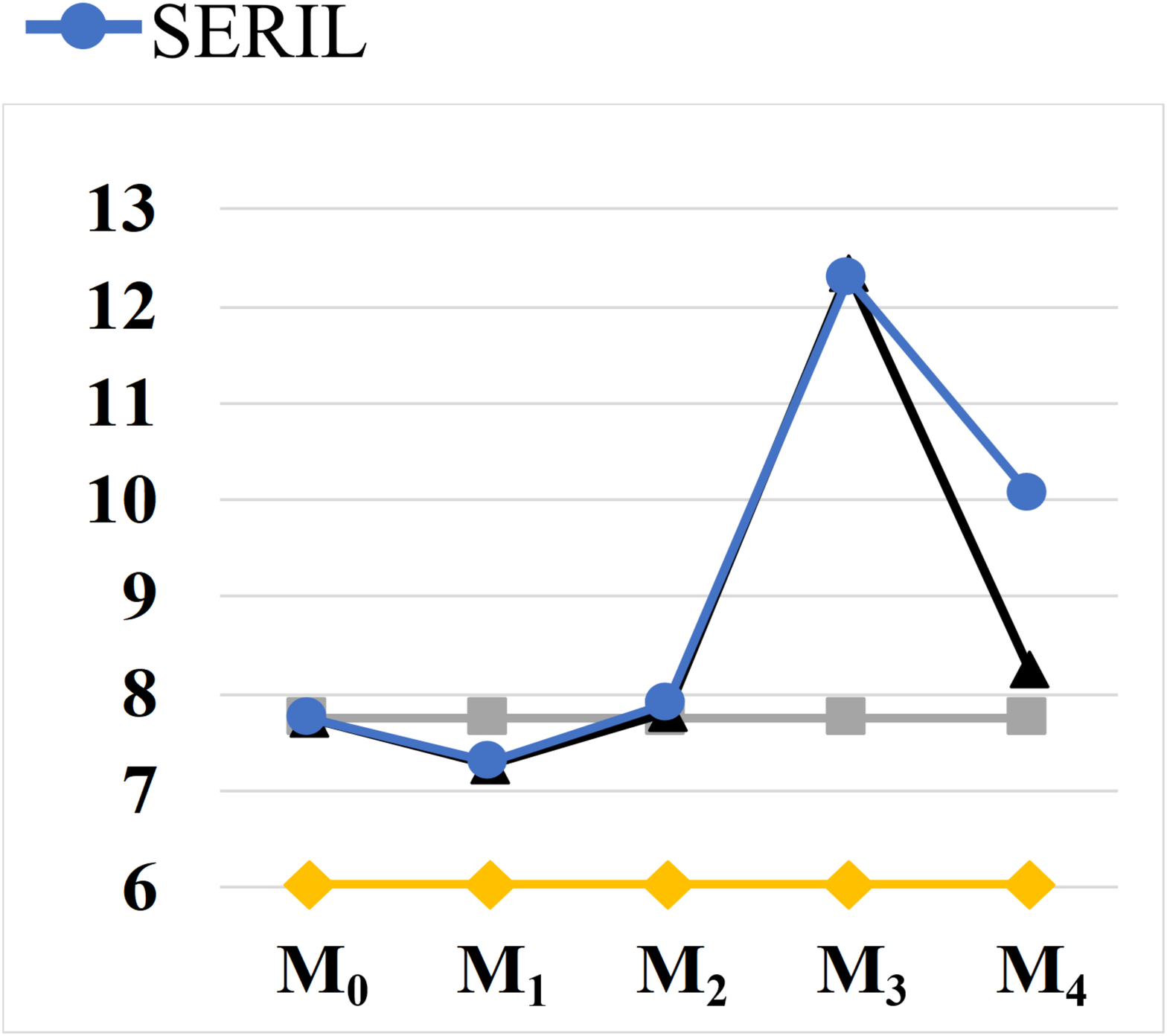}
    }
    \subfigure[$E_{4}$: clap]{
        \includegraphics[width=0.17\linewidth]{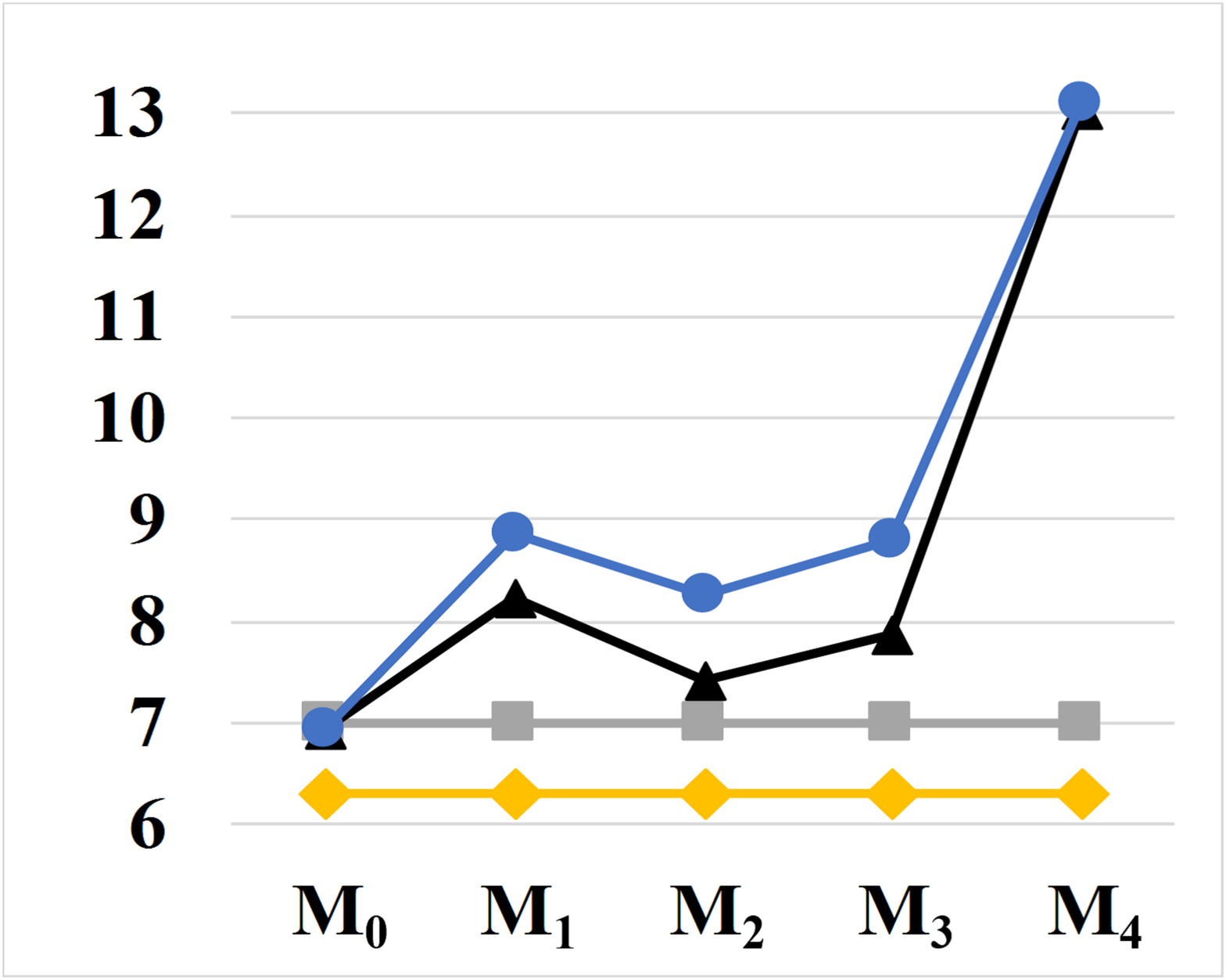}
    }
    \vspace{-0.5em}
    \caption{$SDR^{STSA}$ scores of incrementally learned models evaluated on five testing sets. The x-axis lists incrementally learned models $M_{0}$, $M_{1}$, $M_{2}$, $M_{3}$, and $M_{4}$. The y-axis presents the $SDR^{STSA}$ score. The scores of the unprocessed noisy speech, baseline model, direct fine-tuning approach, and proposed SERIL are represented by yellow, gray, black, and blue lines, respectively.}
    \vspace{-1em}
    \label{fig:exp1}
\end{figure*}

\section{Experiment and Analysis}
\label{sec:evaluation}

\subsection{Experimental Setup}
\label{sec:setup}
We evaluated the proposed SERIL system on two speech corpora: VCB\cite{VCB} and TIMIT \cite{TIMIT}.
Three data sets were prepared, namely, the training, adaptation, and testing sets.
For the training set, 2,000 utterances were randomly selected from the VCB corpus.
Each utterance was contaminated with 13 types of noise (obtained from the NOISEX-92 database\cite{NOISEX-92}) at 6 signal-to-noise (SNR) levels (ranging from -3 dB to 12 dB with a step of 3 dB), amounting to 156,000 (=2000$\times$13$\times$6) paired noisy-clean utterances in total.
This training set is termed $T_{0}$.
To prepare the adaptation sets, we randomly selected another 300 utterances from the VCB corpus.
These 300 utterances were contaminated with other 4 types of noise (obtained from the Nonspeech database \cite{NonSpeech}): \emph{cough}, \emph{door moving}, \emph{footsteps}, and \emph{clap}, at 6 SNR levels (from -3 dB to 12 dB with a step of 3 dB) to form 4 adaptation sets, termed $T_{1}$, $T_{2}$, $T_{3}$, and $T_{4}$.
Each set contained 1,800 (=300$\times$6) paired noisy-clean utterances.

For the testing set, we selected 1,680 utterances from the TIMIT data set.
There were a total of five testing sets.
The first testing set, $E_{0}$, corresponded to the training set $T_{0}$.
The other four testing sets $E_{1}$ to $E_{4}$ corresponded to the adaptation sets $T_{1}$ to $T_{4}$.
For the testing set $E_{0}$, there were 1,680 noisy utterances, and the noise types and SNR levels were the same as those used in $T_{0}$.
Each utterance was contaminated with one of the 13 noise types at a particular SNR level (one out of 6 SNR levels was randomly specified).
For each of the testing sets $E_{1}$ to $E_{4}$, there were also 1,680 noisy utterances, and  each utterance was contaminated with one noise type at a particular SNR level (one out of the 6 SNR levels was randomly specified). Our implementation is publicly available for reproducibility\footnote{\href {https://github.com/ChangLee0903/SERIL}{https://github.com/ChangLee0903/SERIL}}.

Three standardized evaluation metrics were used to measure the performance: perceptual evaluation of speech quality (PESQ) \cite{PESQ}, short-time objective intelligibility measure (STOI) \cite{STOI}, and extended STOI (eSTOI) \cite{eSTOI}.
PESQ was designed to evaluate the quality of processed speech.
The higher the PESQ, the better the speech quality.
Both STOI and eSTOI were designed to compute the speech intelligibility.
The higher STOI and eSTOI scores, the better the speech intelligibility.
In addition, we also reported the $SDR^{STSA}$ scores to illustrate the learning process.
The higher the $SDR^{STSA}$ score, the smaller the distortion of the spectral features.

\subsection{Experimental Results}
\label{sec:result}

First, we compared SERIL and the direct fine-tuning approach in terms of the adaptation capability and the degree of catastrophic forgetting.
We used the training set $T_{0}$ to train one baseline model, termed $M_{0}$.
Then, based on the four adaptation sets, we sequentially adapted the model from $M_{0}$ to $M_{1}$ using $T_{1}$, $M_{1}$ to $M_{2}$ using $T_{2}$, $M_{2}$ to $M_{3}$ using $T_{3}$, and $M_{3}$ to $M_{4}$ using $T_{4}$.
The five models ($M_{0}$ to $M_{4}$) were then tested on the five testing sets ($E_{0}$ to $E_{4}$).
The $SDR^{STSA}$ scores of the five models tested on the five testing sets are shown in Figure~\ref{fig:exp1}.
The results of the baseline model without adaptation and the scores of unprocessed noisy speech are also given for comparison.

From the figure, we note that although the baseline model $M_{0}$ performs well on $E_{0}$, where the noise types and SNR levels are matched during the training and testing stages, notable degradation is observed for the mismatched conditions (cf. the gray lines on $E_{1}$ to $E_{4}$).
Further, both SERIL and the direct fine-tuning approach effectively adapt the SE model to each target domain and achieve good performance.
For example, in Figure~\ref{fig:exp1}(b), $M_{1}$ achieves the best performance on $E_{1}$ for both SERIL and the direct fine-tuning approach.
The model trained by direct fine-tuning tends to forget the previously learned SE capability, whereas the model trained by SERIL can maintain good SE performance for previously learned noise types.
For instance, in Figure~\ref{fig:exp1}(b), the performance of $M_{4}$ trained by direct fine-tuning is considerably reduced in $E_{1}$, showing that the adapted model has ``forgotten'' the SE capability for the previously learned noise type.
This is because each noise type has different structural characteristics in different frequency bands, so direct fine-tuning without proper constraints can severely distort the modeling of previous noise environments.
In contrast, the performance drop of the SERIL system for the same training-testing case is relatively minor.
Consistent trends can be observed for all testing sets.

\begin{table}[ht]
\scriptsize
	\caption{$SDR^{STSA}$, PESQ, STOI, and eSTOI scores of model $M_{4}$ trained by the fine-tuning method (F) and SERIL (R). The results of unprocessed noisy speech (N) and the baseline model $M_{0}$ without adaptation (P) are listed for comparison.}
	\vspace{-0.5em}
	\centering 
	\label{tab:exp2}
	\begin{tabular}{| c | c | c | c | c | c | c |}
	    \hline
		\multirow{2}{*}{Metric} & \multirow{2}{*}{M} & \multirow{2}{*}{original} & \multirow{2}{*}{cough} & door & foot- & \multirow{2}{*}{clap} \\
		& & &  & moving & steps &  \\
		\hline
		
		\hline
		\multirow{4}{*}{$SDR^{STSA}$} & N & 6.23 & 6.43 & 6.87 & 6.05 & 6.31 \\
        & P & \textbf{11.75} & 7.17 & 7.75 & 7.74 & 7.03 \\
        & F & 6.99 & 8.39 & 8.72 & 8.27 & 13.05 \\
        & R & 9.31 & \textbf{10.15} & \textbf{10.97} & \textbf{10.07} & \textbf{13.11} \\
		\hline
        \multirow{4}{*}{PESQ} & N & 2.266 & 2.041 & 1.864 & 1.868 & 1.474 \\
        & P & \textbf{2.708} & 2.118 & 2.059 & 2.015 & 1.603 \\
        & F & 2.406 & 2.204 & 2.339 & 2.133 & \textbf{2.948} \\
        & R & 2.461 & \textbf{2.375} & \textbf{2.581} & \textbf{2.381} & 2.936 \\
		\hline
        \multirow{4}{*}{STOI} & N & 0.816 & 0.788 & 0.743 & 0.778 & 0.789 \\
        & P & \textbf{0.869} & 0.798 & 0.779 & 0.799 & 0.801 \\
        & F & 0.811 & 0.816 & 0.825 & 0.829 & 0.923 \\
        & R & 0.826 & \textbf{0.839} & \textbf{0.859} & \textbf{0.855} & \textbf{0.931} \\
		\hline
        \multirow{4}{*}{eSTOI} & N & 0.624 & 0.692 & 0.648 & 0.744 & 0.782 \\
        & P & \textbf{0.721} & 0.695 & 0.661 & 0.745 & 0.788 \\
        & F & 0.638 & 0.698 & 0.687 & 0.745 & \textbf{0.853} \\
        & R & 0.664 & \textbf{0.717} & \textbf{0.731} & \textbf{0.763} & \textbf{0.853} \\
        \hline
	\end{tabular}
	\vspace{-1.5em}
\end{table}

Table~\ref{tab:exp2} shows the $SDR^{STSA}$, PESQ, STOI, and eSTOI scores of the final model ($M_{4}$) learned using the fine-tuning method and SERIL on the five testing sets.
The scores of unprocessed noisy speech and the baseline model without adaptation ($M_{0}$) are also listed for comparison.
Several observations can be drawn from the table.
First, SERIL performs as well as direct fine-tuning in the current noise environment in terms of all metrics (cf. the ``clap'' column in Table~\ref{tab:exp2}).
Second, SERIL always outperforms direct fine-tuning for previous environments in terms of all metrics (cf. the ``original'' to ``footsteps'' columns in Table~\ref{tab:exp2}).
Third, SERIL performs better than the baseline model in all testing environments except for ``original'', which is under a matched training-testing condition for the baseline model.
It is worth noting that compared with the direct fine-tuning approach, SERIL requires only a small amount of additional computational cost and storage to set the constraints when performing model adaptation.
However, SERIL can produce performance comparable to the direct fine-tuning approach in each new environment while overcoming the catastrophic forgetting problem in old environments.

\section{Concluding Remarks}
\label{sec:conclusion}
When deploying an SE system in real-world applications, it is common to encounter a new noisy environment and re-visit to previous noisy environments.
Although the direct fine-tuning approach can effectively adapt SE models to new environments, the adapted SE model may suffer from the catastrophic forgetting problem.
The proposed SERIL model not only yields comparable performance to the direct fine-tuning approach but also effectively overcomes the catastrophic forgetting problem.
To the best of our knowledge, this paper is the first work that incorporates incremental learning into SE tasks.
Our experimental results confirmed the effectiveness of the proposed SERIL system for SE model adaptation and avoiding catastrophic forgetting.
Based on the promising results, we believe that the proposed SERIL model can be used in various edge-computing devices, where the acoustic condition changes frequently and the cost of online retraining is high.
In addition, we note that using an appropriate weight, $\lambda$, to combine the curvature-based and path optimization-based strategies can provide better SE performance in most tasks.
Derivation of an algorithm that can automatically determine the optimal $\lambda$ is worthy of further study.

\bibliographystyle{IEEEtran}

\bibliography{refs}


\end{document}